\documentclass[twocolumn]{aastex63}
\usepackage{amsmath}
\usepackage{multirow}
\usepackage{CJK}
\turnoffedit

\usepackage[caption=false]{subfig}

\usepackage{graphicx}

\definecolor{darkgreen}{RGB}{30,150,30}
\definecolor{darkblue}{RGB}{0,0,127}

\newcommand{\mulcccit}[1]{\multicolumn{3}{c}{\textit{#1}}}

\newcommand{\mulrccc}[1]{\multirow{3}{*}{\parbox{2.5cm}{\centering #1}}}

\newcommand{\mulrcc}[1]{\multirow{2}{*}{\centering #1}}

\shorttitle{Cosmic-CoNN}
\shortauthors{Xu, McCully, Dong, Howell, and Sen}
\graphicspath{{./}{figures/}}
\begin{document}
\begin{CJK*}{UTF8}{gbsn}

\title{Cosmic-CoNN: A Cosmic Ray Detection \\
Deep-Learning Framework, Dataset, and Toolkit}

\correspondingauthor{Chengyuan Xu}
\email{cxu@ucsb.edu}

\author[0000-0001-9981-8889]{Chengyuan Xu (许程远)}
\affiliation{Media Arts and Technology, University of California, Santa Barbara, CA 93106, USA}
\affiliation{Department of Computer Science, University of California, Santa Barbara, CA 93106, USA}

\author[0000-0001-5807-7893]{Curtis McCully}
\affiliation{Las Cumbres Observatory, 6740 Cortona Drive, Suite 102, Goleta, CA 93117-5575, USA}

\author{Boning Dong (董泊宁)}
\affiliation{Department of Electrical and Computer Engineering, University of California, Santa Barbara, CA 93106, USA}

\author[0000-0003-4253-656X]{D. Andrew Howell}
\affiliation{Las Cumbres Observatory, 6740 Cortona Drive, Suite 102, Goleta, CA 93117-5575, USA}
\affiliation{Department of Physics, University of California, Santa Barbara, CA 93106, USA}

\author[0000-0002-8042-924X]{Pradeep Sen}
\affiliation{Department of Electrical and Computer Engineering, University of California, Santa Barbara, CA 93106, USA}

% \nocollaboration{6}



\begin{abstract}

% A 250 word limit for the abstract. 

% Keep the reader upper-most in your mind. What does the reader know so far? What does the reader expect next and why? 

% Heilmeier Catechism: 1. What are you trying to do? Articulate your objectives using absolutely no jargon. 2. How is it done today, and what are the limits of current practice? 3. What is new in your approach and why do you think it will be successful? 4. Who cares? If you are successful, what difference will it make?

Rejecting cosmic rays (CRs) is essential for the scientific interpretation of CCD-captured data, but detecting CRs in single-exposure images has remained challenging. Conventional CR detectors require experimental parameter tuning for different instruments, and recent deep learning methods only produce instrument-specific models that suffer from performance loss on telescopes not included in the training data. \edit1{We present Cosmic-CoNN, a generic CR detector deployed for 24 telescopes at the Las Cumbres Observatory, which is made possible by the three contributions in this work: 1)~We build a large and diverse ground-based CR dataset leveraging thousands of images from a global telescope network. 2)~We propose a novel loss function and a neural network optimized for telescope imaging data to train generic CR detection models. At 95\% recall, our model achieves a precision of 93.70\% on Las Cumbres imaging data and maintains a consistent performance on new ground-based instruments never used for training. Specifically, the Cosmic-CoNN model trained on the Las Cumbres CR dataset maintains high precisions of 92.03\% and 96.69\% on Gemini GMOS-N/S 1x1 and 2x2 binning images, respectively.} 3)~We build a suite of tools including an interactive CR mask visualization and editing interface, console commands, and Python APIs to make automatic, robust CR detection widely accessible by the community of astronomers. Our dataset, open-source codebase, and trained models are available at \url{https://github.com/cy-xu/cosmic-conn}.

\end{abstract}



\section{Introduction} \label{sec:intro}

\begin{figure*}[ht!]
    \centering
    \includegraphics[width=0.9\textwidth]{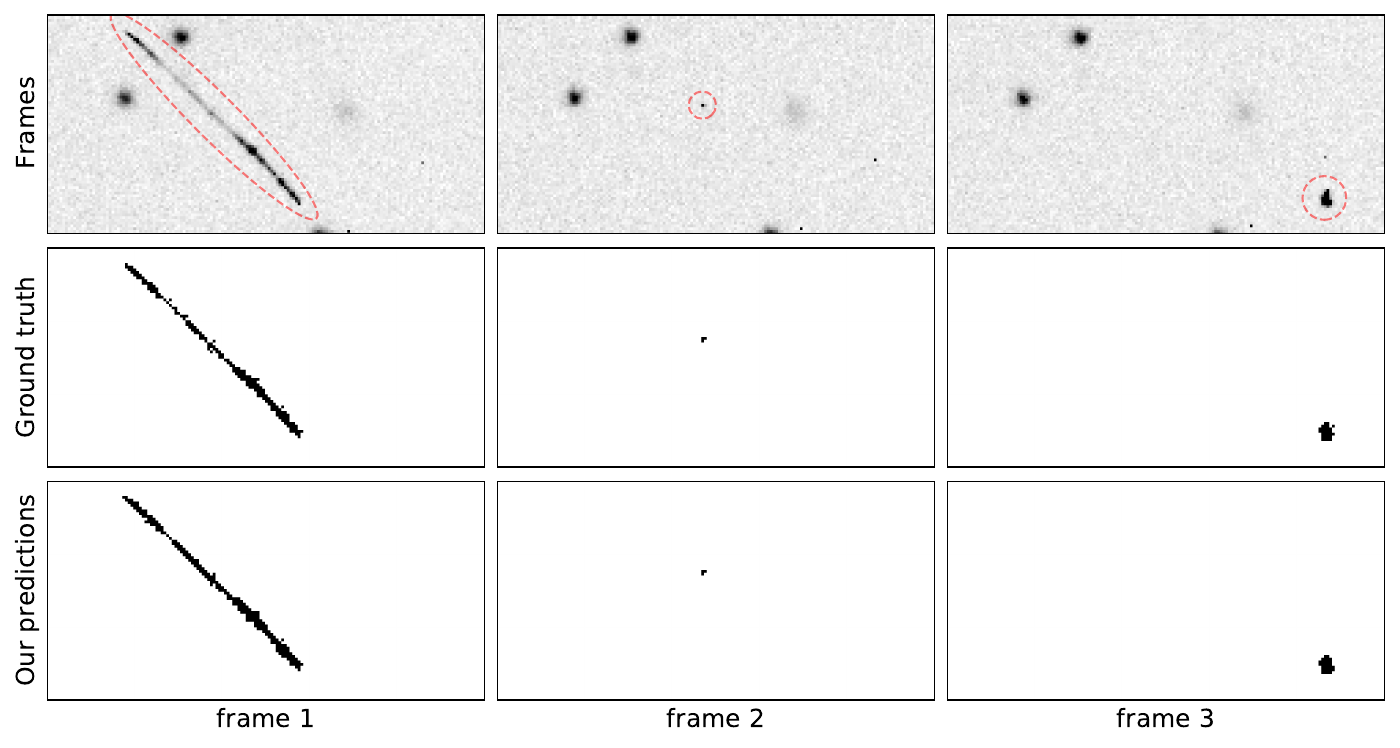}
    \caption{Cosmic rays (CRs), labeled with red circles, and other artifacts can be identified by comparing a pixel's deviation from the pixel location's median value in a stack of aligned exposures taken minutes apart. Our deep-learning model predicts a probability map $P$ where $P_{ij}\in [0, 1]$ indicating the likelihood of a pixel being affected by CR using a single frame.}
    \label{fig:cr_triplets}
\vspace*{5pt}
\end{figure*}

Cosmic rays (CRs) are a key source of artifacts in data from astronomical observations using charge-coupled devices (CCDs). These charged particles excite electrons in the detector, creating artifacts that can be mistaken for astronomical sources. Space-based instruments like the \textit{Hubble Space Telescope} (\textit{HST}), which are not protected by Earth's atmosphere, are heavily affected by CR, with an average flux density of $0.96$ $CR/s/cm^2$ \citep{2020arXiv200600909M}. Ground-based instruments are also affected but at a rate about five orders of magnitude lower, typically of $\sim0.00001$ $CR/s/cm^2$ in thin CCDs, as observed in \textit{Las Cumbres Observatory (LCO)} global telescope network imaging data. CCD thickness is another factor that affects an imager's sensitivity to CRs.

% In 4518 LCO images, average CR flux = 1.114182518399658e-05 CR/s/cm2, CR energy = 0.04934813403274722 e/s/cm2

Detecting CRs is straightforward when multiple exposures of the same field are available (see example in Fig.~\ref{fig:cr_triplets}). By comparing the deviation of a pixel from the mean or median value in a stack of aligned images, CRs (and other artifacts) can be effectively identified \citep{1994PASP..106..798W, 1995ASPC...77..514Z, 1995PASP..107...85F, 2002PASP..114..144F, 2016A&C....16...67D}. However, multiple exposures may not be available, especially for spectroscopic observations. Variations in image quality (e.g., seeing) can also complicate this procedure, so robust detection of CR pixels on individual images is still necessary.

CRs do not travel through the telescope's optical path nor do they follow the point spread function (PSF): they are not blurred by the atmosphere and are therefore sharper than a real PSF. Furthermore, they can come in any incidence angle to have less symmetrical morphologies than real astronomical sources. Several algorithms leverage this feature, like adapted PSF convolution \citep{2000PASP..112..703R}, histogram analysis \citep{2004PASP..116..148P}, fuzzy logic-based algorithms \citep{2005AN....326..428S}, and Laplacian edge detection \citep{2001PASP..113.1420V}. These methods and the IRAF task like \texttt{xzap} by M. Dickinson often require adjusting one or more hyper-parameters experimentally to obtain the best result per image. Machine learning algorithms like k-nearest neighbors, multilayer perceptrons~\citep{1991ESOC...38...51M}, and decision-tree classifiers~\citep{1995PASP..107..279S} showed promising results on small HST datasets, but lacked generality when compared to image-filtering techniques like \texttt{LA Cosmic} \citep{2001PASP..113.1420V}.

Machine-learning methods have been widely adopted in astronomical research recently (see \cite{2019arXiv190407248B} for a review). \cite{2020ApJ...889...24Z} used a convolutional neural network (CNN) to identify CR contaminated pixels in \textit{Hubble Space Telescope (HST) ACS/WFC} images, in a method called \texttt{deepCR}. In contrast to using the Laplacian kernel \citep{4767946} for edge detection as is in LA Cosmic, CNNs learn the intrinsic characteristics of the CR artifacts, enabling it to detect CRs of arbitrary shapes and sizes.

The deepCR model outperforms the state-of-the-art method LA Cosmic without manual parameter tuning, demonstrating the promise of deep learning for CR detection. However, its neural network architecture is an adaptation from U-Net \citep{2015arXiv150504597R} which was originally designed for biomedical images, limiting its ability to train a generic model for astronomical observations from different instruments, specifically ground-based data with variable conditions from multiple instruments. \edit1{Furthermore, the low CR rates in ground-based data: a $\sim$1:10,000 ratio between CR and non-CR pixels leads to an extreme class-imbalance issue \citep{BUDA2018249} that provides too few CR pixels for spatial convolution, rendering the training on LCO data more difficult comparing to HST data.}

% Furthermore, the low CR rates in ground-based data makes it more difficult to train a CNN-based model. The $\sim$1:10,000 ratio between CR and non-CR pixels in our ground-based training data leads to the class-imbalance issue \citep{BUDA2018249} that provides too few CR pixels for spatial convolution, rendering the learning very inefficient.

\begin{deluxetable*}{crcccccc}[ht!]
    \tabletypesize{\footnotesize}
    % \tabletypesize{\scriptsize}
    \tablecaption{LCO science imagers covered in the CR dataset.
    \label{table:lco_imagers}}
    \tablehead{
        \colhead{Imager}  \vspace{-0.3cm} & \colhead{Class} & \colhead{Pixel Scale} & \colhead{Binning} & \colhead{Format} & \colhead{Pixel Size} & \colhead{FOV} & \colhead{Filters} \\
        & & \colhead{($''$)} & & \colhead{(pixels)} & \colhead{(microns)} & \colhead{($'$)} &
    }
    \startdata
        SBIG 6303 & 0.4 m & 0.571 & 1$\times$1 & $3K\times2K$ & 9 & $29\times29$ & 9 \\
        Sinistro & 1 m & 0.389 & 1$\times$1 & $4K\times4K$ & 15 & $26\times26$ & 21 \\
        Spectral & 2 m & 0.304 & 2$\times$2 & $4K\times4K$ & 15 & $10\times10$ & 18 \\
    \enddata
\end{deluxetable*}

To address these issues, we present Cosmic-CoNN, a deep-learning framework designed to train generic CR detection models for ground-based instruments by explicitly addressing the class-imbalance issue and optimizing the neural network for the astronomical images' unique spatial and numerical features. Cosmic-CoNN also generalizes to other types of data like space-based and spectroscopic observations.

We leverage the publicly available data from \textit{Las Cumbres Observatory (LCO)} to build a large, diverse CR dataset. \textit{LCO}'s \texttt{BANZAI} data pipeline \citep{2018SPIE10707E..0KM} ensures data from different telescopes is not dominated by instrumental signature artifacts. It allows us to label CRs consistently in thousands of observations taken across a wide variety of sites with diverse scientific goals. The LCO CR dataset promises the rich feature coverage required for a generic CR detection model that would work for a variety of ground-based instruments.

This paper is organized as follows: we present the LCO CR dataset in \S\ref{sec:dataset} and discuss the deep-learning CR-detection framework in \S\ref{sec:dl_framework}. Extensive evaluations on various types of observations are presented in \S\ref{sec:results}. We introduce the toolkit and the software APIs in \S\ref{sec:toolkit}, and conclude the paper with a discussion in \S\ref{sec:conclusion}.

\section{LCO CR Dataset} \label{sec:dataset}

Deep-learning models are data driven. A robust and generic CR-detection model requires a large number of diverse observations from various instruments and the CRs need to be labeled accurately and consistently across different instruments. With this in mind, we build a custom Python CR-labeling pipeline to generate a large cosmic ray ground-truth dataset, leveraging some unique characteristics of \textit{Las Cumbres Observatory} (\textit{LCO}) global telescope network.

% evidence of previous dataset?

Our CR-labeling pipeline stacks consecutive images of the same field to identify cosmic rays. To limit artifacts due to variations in CCD response, we only selected sequences that have at least three repeated observations with  \edit1{identical exposure time and filter}. The LCO CR dataset consists of over 4,500 scientific images from \textit{LCO}'s 23 globally distributed telescopes. About half of the images are $4K\times4K$ pixels resolution and the rest are $3K\times2K$ or $2K\times2K$. To the best of our knowledge, this is the largest cosmic ray dataset that identifies CRs in science images across various ground-based instruments. Each sample in our dataset is a multi-extension \texttt{FITS} file including three images, the corresponding CR masks, and ignore masks. We encoded hot pixels, pixels with no data, and astronomical sources in the ignore masks to reject false-positive CR pixels. The implementation of our ground-truth CR-labeling pipeline is presented in Appendix~\ref{appendix:cr_pipeline}. The LCO CR dataset is available for download at \href{https://zenodo.org/record/5034763}{\tt https://zenodo.org/record/5034763}.

% is key to allow?

\edit1{The dataset covers a} variety of CCD imagers with different pixel scales, field of views, and filters used in \textit{LCO}'s global telescopes network (Table~\ref{table:lco_imagers}). From a deep-learning perspective, diverse data greatly benefits model generality. But having ground-truth CRs labeled consistently on different instruments is not a trivial task. The \texttt{BANZAI} data reduction pipeline \citep{2018SPIE10707E..0KM} performed instrumental signature removal (bad-pixel masking, bias and dark removal, flat-field correction), making \textit{LCO} data suitable for building such a dataset. Instrument artifacts exist as two identical CCDs could have different response curves after years of bombardment by photons and cosmic rays. The standardized data reduction is key to allow our CR-labeling pipeline to consistently and accurately label CRs across various instruments.

\begin{figure}[t]
    % \vspace{-1.2cm}
    \centering
    \includegraphics[width=0.48\textwidth]{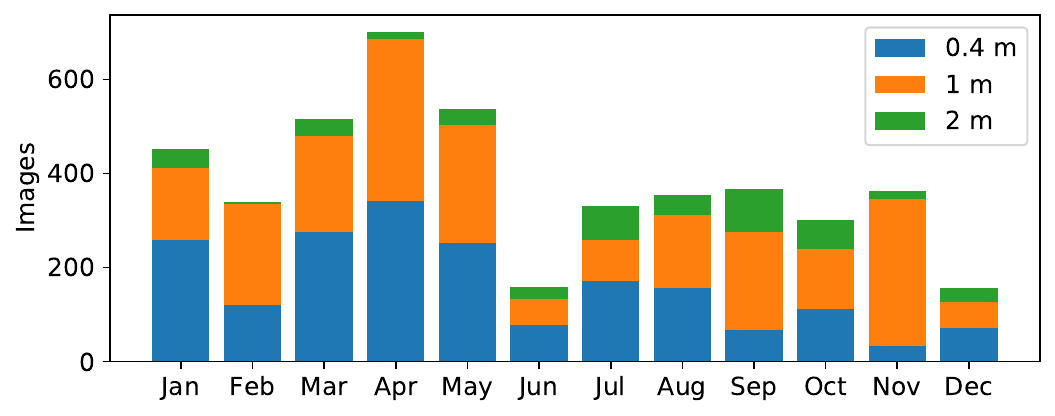}
    \caption{LCO CR dataset sample distribution by month and telescope class, from November 2018 to December 2019. Diverse source densities sampled around the year help improve model robustness.}
    \label{fig:data_stats}
\end{figure}

We chose images from across three telescope classes and across the year as shown in Fig.~\ref{fig:data_stats}. Images from different times of the year sampled a variety of source densities for different sets of scientific goals. The varying source density proved to be of great importance to robust CR detection~\citep{2005PASA...22..249F}. In the task of CR detection, diversified real objects provide rich features for the negative class, which greatly improves model robustness.

\begin{figure*}[ht!]
    \includegraphics[width=1.0\textwidth]{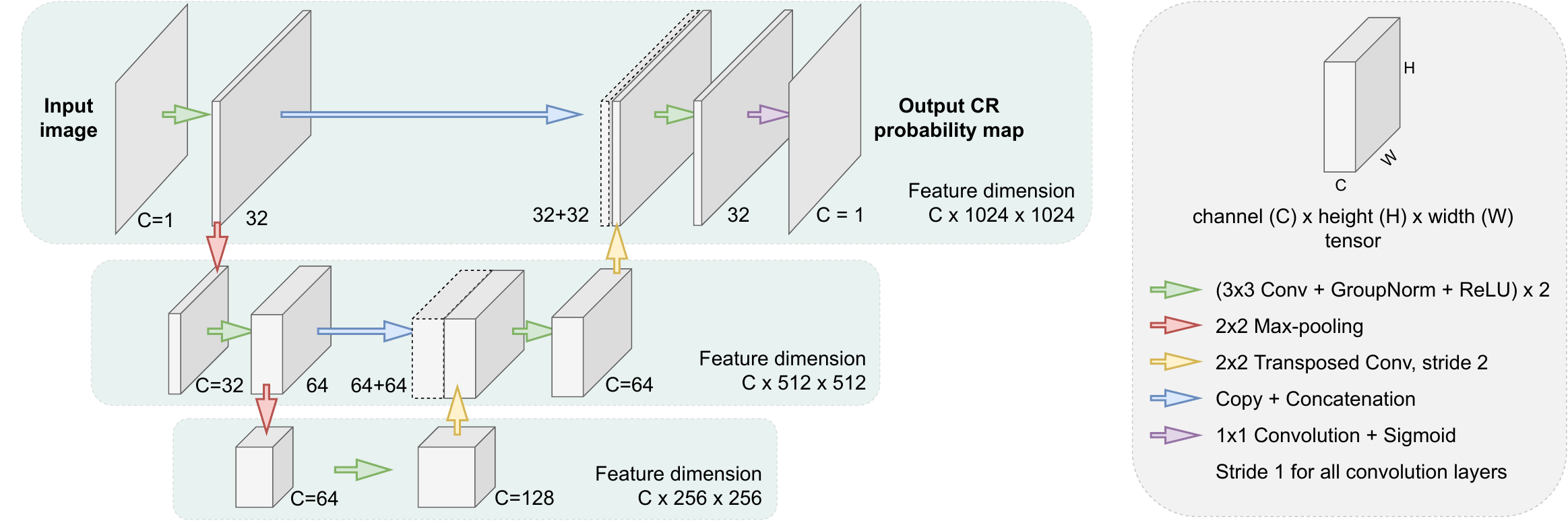}
    \caption{Cosmic-CoNN's neural network architecture is based on U-Net. The symmetric design concatenates high-resolution features from the downsampling path to the upsampling path via skip connections (blue arrows), \edit1{allowing the network to propagate contextual information to higher resolution layers, thereby producing pixel-level classification predictions on CRs of arbitrary shapes and sizes.}}
    \label{fig:model}
    \vspace{0.3cm}
\end{figure*}

We further constrained a sequence of exposures to come from the same scheduling unit: the frames are typically separated by just a few minutes. Repeated exposures in a short period of time help mitigate the PSF variation induced by atmospheric attenuation but PSF wings still cause noticeable false positive labels adjacent sources. We reject CRs that are overlapping with astronomical sources so that variations in the PSF do not create artifacts in the training samples.

Of all CR pixels, 1.21\% were rejected in an effort to tackle the PSF-variation-induced artifacts. This trade-off ensures the remaining 98.79\% CR pixels are labeled at higher confidence. Therefore, models trained with this dataset focus on distinguishing CRs from real sources, and it is anticipated that CRs overlapped with sources will not be detected. Training on raw images with arbitrary PSFs also guarantees consistent performance at inference time. In future versions we will model the PSF explicitly to make sure that we do not bias our training sample.

% About 0.012091580940965116 (mean) 0.008085776390098815 (median) of total CRs (pixels) affected by ignore mask.

Our dataset is not affected by transient sources that evolve at a timescale of hours or longer because of the very tight space between exposures. At this timescale, near-Earth objects (NEOs), satellites, and airplanes could still cause false-positive labels in the stack-based CR masks. Large satellite or airplane trails are rejected by our CR-labeling pipeline automatically. A very small fraction of false-positive labels from NEOs and satellites exist but we have manually verified every single mask to ensure their impact is negligible.

\section{Deep-learning framework} \label{sec:dl_framework}

% The \texttt{U-Net} is a CNN-based deep-learning architecture which takes an image as input and outputs a probability map of the same size . \texttt{U-Net}  and concatenates features of the same scale with skip connections, allowing the network to propagate contextual information to higher resolution layers, thereby producing pixel-level classification predictions on large images.

\edit1{Cosmic-CoNN's neural network architecture} is inspired by the recent success of deepCR \citep{2020ApJ...889...24Z}, a U-Net \edit1{\citep{2015arXiv150504597R}} based deep-learning framework that identifies CR-contaminated pixels in imaging data. \edit1{In contrast to the unique Laplacian kernel used in LA~Cosmic \citep{2001PASP..113.1420V}, a deep CNN model optimizes millions of kernel parameters during training and outputs a pixel-level probability map directly. \edit1{The U-shaped architecture (Fig.\ref{fig:model}) convolves the image at multiple scales},} creating a larger receptive field in deeper layers of its hierarchical architecture to capture not only CRs' morphological features (edges, corners, or sharpness) but also the contextual features from peripheral pixels, allowing it to predict CRs of arbitrary shapes and sizes.

% reported $75.2\%$--$93.3\%$ true-positive rates at a false-positive rate of 0.05\% on different types of \textit{HST ACS/WFC} observations. It

\edit1{deepCR demonstrates the promise of using CNN-based model for CR detection on \textit{HST ACS/WFC} observations.} However, training on ground-based images exposes a number of network architecture and data-sampling limitations it inherited from the U-Net \citep{2015arXiv150504597R}. First, it is worth noting that U-Net was initially proposed to solve biomedical image segmentation problems. The higher dynamic range and extreme spatial variations found in astronomical images need to be addressed explicitly in order to optimize the neural network for these special features in astronomical data. In addition, the high CR rates in \textit{HST ACS/WFC} data does not reflect the extreme class-imbalance issue observed in \textit{LCO} imaging data. The low CR rates make it difficult for deepCR to train and converge on the ground-based LCO imaging data.

In deepCR, \cite{2020ApJ...889...24Z} adopted a two-phase training design to address some of these issues. Assuming correct data statistics are learned in the initial phase, the model freezes feature normalization parameters in the second phase in order to converge. This design works when the inference data shares the same statistics with training data, i.e., an instrument-specific model could be learned. But it works against our goal of a generic CR detection model that works for a wide variety of ground-based instruments with varying data statistics.

Cosmic-CoNN adopted the U-shaped architecture and proposed: (\S\ref{subsec:loss}) a novel loss function that specifically addresses the class-imbalance issue, and (\S\ref{subsec:sampling}) adopted data sampling, augmentation, and feature normalization approaches that are more suitable for ground-based data that work jointly to improve model generality and training efficiency. 

\subsection{Median-weighted loss function} \label{subsec:loss}

The CR-detection task is in essence a pixel-wise binary classification problem. Our goal is to learn a function $f$ which takes an image $I$ as input and outputs $P$, the probability map of each pixel being affected by CR:\\ $P = f(I), P_{ij} \in \left[0, 1\right]$, where $ij$ is the pixel coordinate. The user could then apply an appropriate threshold on $P$ to acquire the binary CR mask. 

Binary cross entropy (BCE) is commonly used to optimize classification models, which can also be used to calculate the loss between the prediction $P$ and the ground-truth CR mask $Y$:
\begin{equation}
    \begin{aligned}
        \label{eq:bce}
        \text{BCE}(P, Y) = - & (Y_{ij}\log(P_{ij}) + \\
                          & (1 - Y_{ij})\log (1 - P_{ij})) &&
    \end{aligned}
\end{equation}
where the ground-truth mask $Y$ is defined as $Y_{ij}=1 $ for CR pixels and $Y_{ij}=0$ for non-CR pixels. The first term $Y_{ij}\log(P_{ij})$ measures the loss for CR pixels and second term for non-CR pixels. The optimization objective is to minimize their sum to account for both CR and non-CR classes.

The low CR rates in LCO data causes the non-CR loss to dominate the total loss. Training on \textit{LCO} imaging data, the observed losses from the two terms in Equation~\ref{eq:bce} have a ratio of $\sim$1:6300 (averaged over 10 random experiments), with the second term (non-CR loss) dominating the optimization objective. This verifies the class-balance issue. 

% Therefore, we make a hypothesis that the network will rapidly learn an incorrect function which predicts $0$ (not CR) for all pixels since this is the quickest path to minimize the total loss. Close observation to the training process proved our hypothesis correct: regardless of the initialization, the model trained with BCE loss predicts $\sim$0.0 for all pixels in just a few iterations, i.e.,\ the incorrect function is learned because loss from the second term dominates the optimization objective.

\begin{figure}[t]
    \includegraphics[trim={8 0 0 0},clip,width=0.48\textwidth]{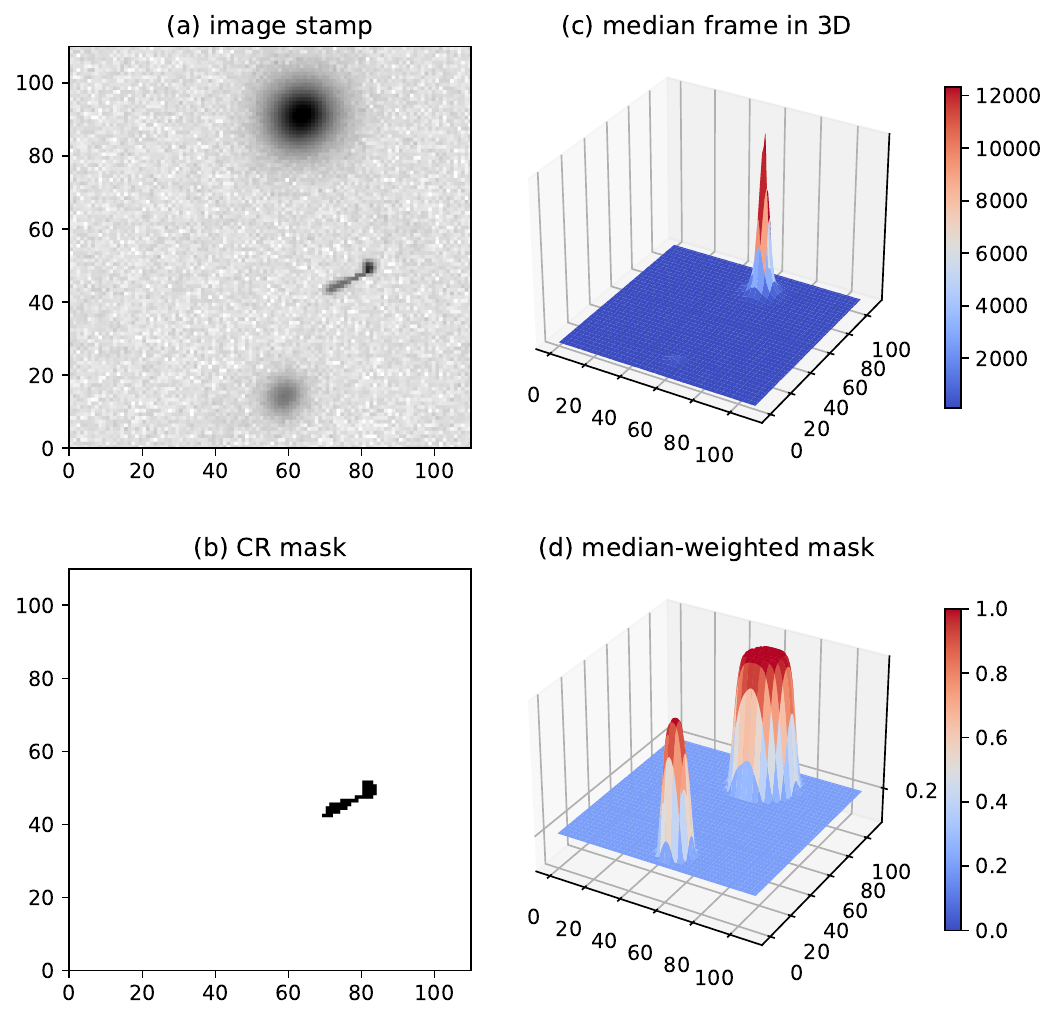}
    \caption{3D visualization of the median-weighted mask. \edit1{(a) An image stamp that includes sources, CR affected pixels, and background. (b) The ground-truth CR mask shows the imbalance between CR and non-CR pixels. (c) 3D visualization of the CR-free median image shows the non-CR pixels can be further split into two sub-classes: sources and background, while the background pixels may be dominant in quantity. We transform (b) to acquire (d), the median-weighted mask $(M)$ by normalizing the brightness variation between sources. $M$ in Eq.~\ref{eq:median_bce} adaptively down-weight background pixel loss in the proposed median-weighted loss function.} In this figure, $M_{ij} \in \left[0.2, 1.0\right]$.} 
    \label{fig:median_mask}
\end{figure}

Furthermore, background pixels are the culprit for an extra layer of imbalance within the non-CR class. From dark background to bright sources, the non-CR class often covers the image's entire dynamic range (see example in Fig.~\ref{fig:median_mask}a,b). Although both labeled as $0$ in $Y$ (Fig.~\ref{fig:median_mask}c), the lopsided numerical difference between background and sources in fact creates two sub-classes within the non-CR class to introduce inconsistency, making the training path even more convoluted. 

% Close observation to the training process proved us right again: followed by learning the aforementioned incorrect functional that predicts $\sim$0.0 for all pixels, we observe the model's prediction for both source and CR pixels approaching 1.0 (CR) simultaneously, i.e., as the model revises itself from the first incorrect function, it learns a second incorrect function that tend to predict all bright objects as CRs.

The class imbalance and the numerical imbalance within the non-CR class are clear indications that we should directly focus on learning to distinguish between CRs and sources. It inspired us to create an adaptive per-pixel weighting factor that prioritizes on CR and source pixels by down-weighting the less useful yet dominant loss from background pixels.
% By avoiding the inefficient and convoluted learning path, the model could then converge to a better minimum.

%  the brightness variation between sources makes it hard to use the median frame as a weight mask directly. 

Since we already \edit1{acquired} a sequence of consecutive exposures building the LCO CR dataset, we could use the CR-free median frame (Fig.~\ref{fig:median_mask}b) as an unique ground-truth to separate sources from the background. The brightness variation between different sources makes it hard to use the median frame as a weight mask directly, so we perform a series of transformations (sky subtraction, clipping between one and five robust standard deviations, $5\times5$ kernel with $\sigma=2$ Gaussian smoothing, unit normalization, and finally clamping with a lower-bound parameter $\alpha$) to separate sources from the background to acquire the median-weighted mask (M) shown in Fig.~\ref{fig:median_mask}d. We apply $M$ to the non-CR loss term in BCE to get the novel median-weighted loss function $(L_{M})$:
\begin{equation}
    \begin{aligned}
        \label{eq:median_bce}
        L_{M}(P, Y, M) = - & (Y_{ij}\log(P_{ij}) + \\
                          & \mathbf{M_{ij}}(1 - Y_{ij})\log (1 - P_{ij})) &&
    \end{aligned}
\end{equation}
where $M_{ij} \in \left[\alpha, 1\right]$. Pixel by pixel, $M$ adaptively down-weights the loss from background by scaling with the lower bound $\alpha$, mitigating the extreme imbalance between the two loss terms and redefines the optimization objective to directly learning to distinguish between sources and CRs.

\begin{figure*}[t]
    \includegraphics[width=1\textwidth]{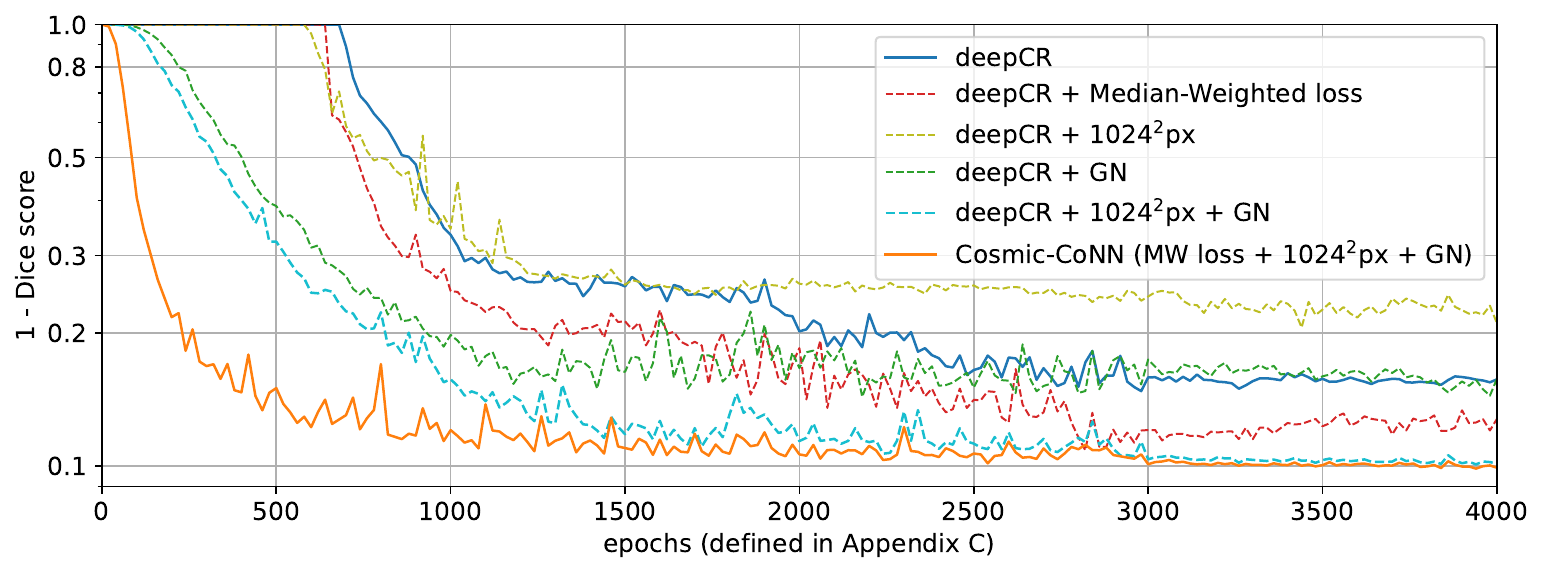}
    \caption{Using deepCR as baseline, we demonstrate our proposed improvements' effects to the model performance as a function of training progress. All variant models are initialized with the same random seed, trained on an identical set of LCO data, and evaluated with identical validation images using same model-input dimension. Performance is measured by the S{\o}rensen{-}Dice coefficient \citep{sorensen1948method} (henceforth, the Dice score) to gauge the similarity between the model's prediction and the ground-truth CR mask. Here we plot (1 - Dice score) in logarithmic scale, lower is better. Models without using group-normalization (GN) were trained in two phases, thus the delayed optimizations that start after 500 epochs. The median-weighted loss help deepCR to achieve better performance, while the larger $1024^2$ pixels stamps proved to be vital for models using GN. The proposed median-weighted loss function, increased stamp size, and GN work jointly to allow Cosmic-CoNN to converge rapidly and to a better minimum. Quantitative results are presented in Table.~\ref{table:ablation} in ablation study (Appendix \ref{sec:ablation_study}).}
    \label{fig:training_evaluation}
    \vspace{0.3cm}
\end{figure*}

With $M$ applied to the second term in BCE, it immediately reduces the observed CR to non-CR class losses to $\sim$1:300 in Equation~\ref{eq:median_bce}, comparing to the $\sim$1:6300 using Equation~\ref{eq:bce} (in identical conditions). Although this ratio can be further reduced with a more aggressive weight mask, the median-weighted mask preserves all real sources without introducing inconsistency. After training with 500 images, the observed loss of the two terms further reduce to $\sim$1:6 using $L_{M}$, comparing to $\sim$1:110 using BCE loss. In Fig.~\ref{fig:training_evaluation}, we show that the deepCR model optimizes sooner and to a better minimum with $L_{M}$ while holding other variables constant. 

The median-weighted loss function ($L_{M}$) makes use of the median frame's unique CR-free property as a robust weighting factor to effectively suppresses the dominating loss from background pixels, at the same time prioritizes on learning to distinguish between CRs and sources by maintaining their weighting factor at $1.0$. As training progresses, the lower bound $\alpha$ linearly increases the weight for background pixels from $0.0$ to $1.0$ so the model could learn a clear boundary for CRs. 

We could also cap $\alpha$ at less than 1 to learn a model that produces CR prediction with soft edges, leaving more control to the user-defined threshold when a binary CR mask is needed. We choose to increase $\alpha$ to~1 so that $L_{M}$ converges to the BCE loss, working with the standard Sigmoid function \citep{LITTLE1974101, LITTLE1978281} at the last layer of our network to produce a theoretical best classification boundary of around 0.5. We also experimented using a loss function based on S{\o}rensen{-}Dice coefficient that is robust for imbalanced data \citep{7785132} but the model learned a strong bias to avoid CRs near real objects, making the more interpretable BCE-based loss a better choice for optimization. 

% Our novel Median-Weighted loss function ($L_{M}$) makes use of the median frame's unique CR-free property as a robust weighting factor to guide the model's learning path. $L_{M}$ converges the model to a better minimum by preventing the model from learning incorrect functions during training, and also improves training efficiency with a better convergence path.

% move title to next column
% \newpage

\subsection{Data sampling and normalization} \label{subsec:sampling}

% Current observations are often saturated in the analog/digital conversion (ADC) process but future instruments might require even higher bit-depth and become more challenging for existing model designs (to handle gradient diminishing or gradient explosion issues).

Large-scale deep-learning models are often optimized using stochastic gradient descent \citep{kiefer1952stochastic}, motivated by stochastic methods' efficiency benefits, at the same time constrained by the ever-growing dataset size and limited GPU memory (usually on the order of 10 GB) for parallel computation. Model parameters are iteratively optimized over a small batch of data, colloquially known as a mini-batch, randomly sampled from the full dataset. If iterating over all $N$ samples in a dataset is considered an \textit{epoch}, then training a model with $n$ samples in a mini-batch means the model updates about $ \lfloor\frac{N}{n}\rfloor$ times in an epoch \citep{2016arXiv160604838B}.

\edit1{By slicing \textit{HST ACS/WFC} images into $256^2$ pixel stamps, deepCR \citep{2020ApJ...889...24Z} samples a mini-batch from a dataset of fixed stamps. However, this approach} is unsuitable for ground-based astronomical images featuring much lower CR rates: a small $256^2$ stamp might not include a single CR, making many of the samples less useful for training. 

% Because CNN-based models extract features and optimize kernel weights through spatial convolution, and Equation~\ref{eq:median_bce} shows our loss function optimize the model using both CR and non-CR features. 

% explain what is a decent count

Recall that each sample in the LCO CR dataset is a multi-extension \texttt{FITS} including three images between $2K\times2K$ and $4K\times4K$ pixels. This design empowers a more flexible data-sampling strategy than having the dataset stored in a fixed size. The Cosmic-CoNN framework could crop a stamp of any size, up to the entire image from each \texttt{FITS}, ensuring a reasonable number of CRs in every mini-batch. The sparsity of source and CR in ground-based astronomical data motivated us to increase the sampling stamp size to $1024^2$ pixels. A larger area is more likely to include all three types of features: sources, CRs, and background in a single stamp and also provides more spatial and contextual information for the convolution operations in CNN models.

One consequence of the increased stamp size is the decreased number of samples in a mini-batch, given the same amount of GPU memory. Increasing the stamp width and height by $m$ times will reduce the batch size $n$ to $\lfloor\frac{n}{m^2}\rfloor$, e.g., the memory that fits a mini-batch of $16\times256^2$ pixel images can only fit a single $1024^2$ pixel image. The accuracy of batch normalization (BN) \citep{DBLP:conf/icml/IoffeS15}, an important feature-normalization method widely used in deep CNN architectures, including in deepCR, decreases rapidly when the batch size becomes too small, so adopting the proposed larger stamp size alone might even hurt model accuracy, as shown in Fig.~\ref{fig:training_evaluation}. We adopt group normalization (GN) \citep{DBLP:journals/corr/abs-1803-08494}, whose computation is independent of batch size to address the accuracy loss in BN. Unlike BN which normalizes over all feature channels across all samples in a mini-batch, GN divides feature channels into groups and computes the normalization statistics for each sample. We used GN as a remedy for the decreased batch size but found it playing a major role in improving training efficiency on astronomical imaging data.

The high dynamic range, high variance, low source density, and low CR rates in ground-based astronomical images make it difficult to learn accurate per-sample normalization statistics from small stamps: one sample could include a bright source but another could be entirely dark. By pairing GN with the proposed stamp size of $1024^2$ pixels, the learned per-sample normalization is more accurate because of the extra spatial and contextual information from the wider field of view.

As a common practice in deep-learning research, we conduct an ablation study to demonstrate the individual and combined effects of median-weighted loss, $1024^2$px sampling size, and GN. The results are presented in Fig.~\ref{fig:training_evaluation} and Appendix~\ref{sec:ablation_study}. Controlled experiments show applying GN alone improves training efficiency but not model performance. By pairing GN with the increased $1024^2$ stamps, it dramatically improves performance and model generality, while the proposed new loss function provides Cosmic-CoNN a better convergence path to further improve the model's performance and generality on both \textit{LCO} and \textit{Gemini} instruments (see Table.~\ref{table:ablation}).

Finally, in addition to randomly cropping image stamps form a large image, we perform weak data augmentation like random rotations as well as horizontal and vertical mirroring, allowing the model to learn invariance to pose variation in astronomical observations \citep{2018A&C....25..103G}. Strong augmentations like elastic deformations adopted by \cite{2015arXiv150504597R} have proved to be effective to improve performance on a small dataset but we avoided such deformation as it could change real CRs' sharp profiles. Given the large number of diverse samples in LCO CR dataset, we found weak augmentations sufficient. With pose augmentation, we also saw more stabilized training and improved performance on \textit{HST ACS/WFC} data, showing that weak augmentation is effective in increasing model robustness.

\begin{figure*}[t!]
    \centering
    \includegraphics[clip,width=0.95\textwidth]{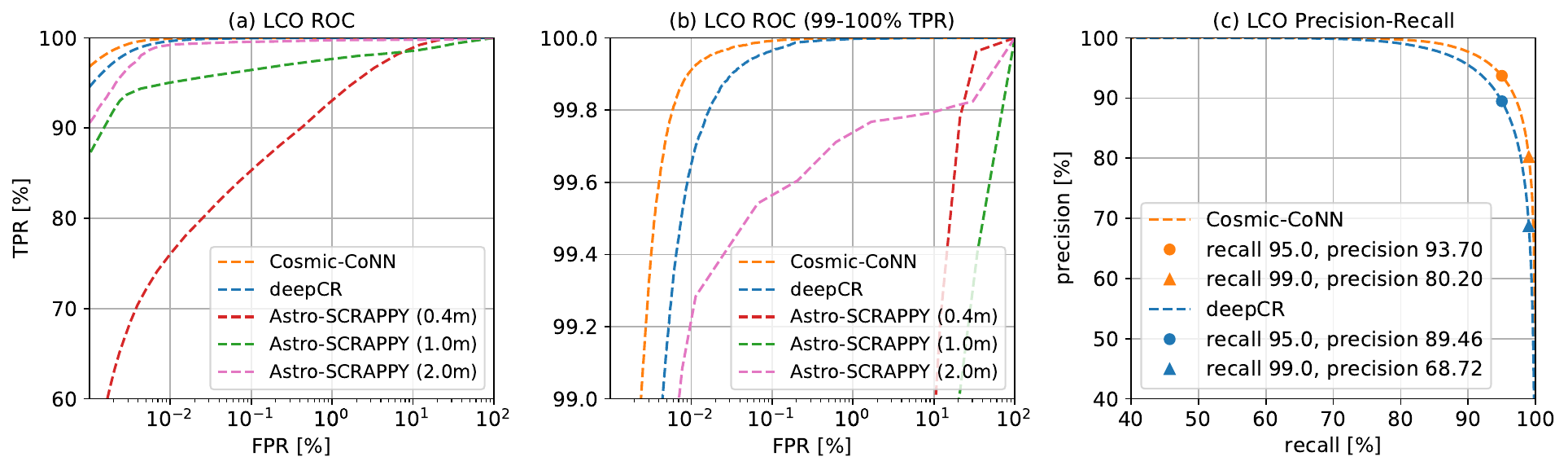}
    \caption{Evaluating three CR detectors with ROC and Precision-Recall curves on \textit{LCO} imaging data. It is desirable to have a higher true-positive rate (TPR) at fixed false-positive rates (FPR) in ROC (Equation~\ref{eq:roc_tpr},\ref{eq:roc_fpr}). As illustrated in (a) and (b), Cosmic-CoNN outperforms other methods with higher TPRs overall. The margin of its lead further increases in more strict low FPRs, showing Cosmic-CoNN's robust performance. \edit1{Circle} markers on the Precision-Recall curves in (c) show when $95\%$ of the CR pixels are found (\edit1{95\%} recall), Cosmic-CoNN's prediction is over $4\%$ more accurate than deepCR (precision). At $99\%$ recall, Cosmic-CoNN's lead increases to $\sim$$11\%$.}
    \label{fig:lco_roc_pr}
\end{figure*}

\begin{figure*}[t!]
    \centering
    \includegraphics[trim={0 0 0 6},clip,width=0.95\textwidth]{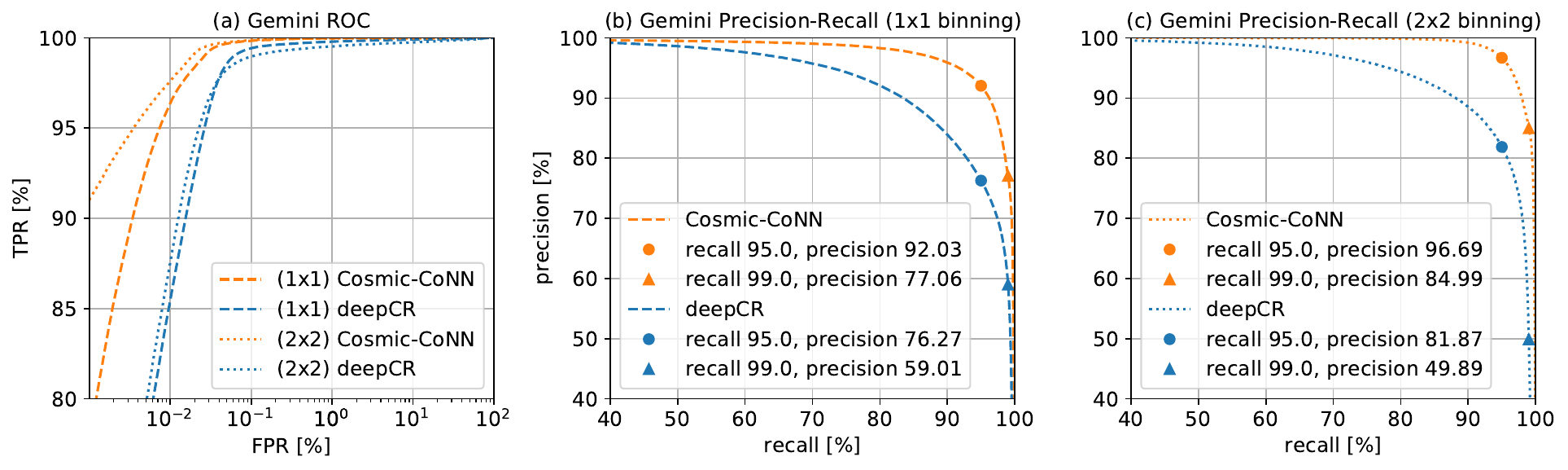}
    \caption{\edit1{We trained both deepCR and Cosmic-CoNN on LCO data and evaluate their performance on new images from previously unseen Gemini GMOS-N/S telescopes. Comparing Gemini data's Precision-Recall curves (Fig.7bc) with LCO's (Fig.6c) shows the Cosmic-CoNN model maintains similar performance while deepCR has visible performance loss (see precision gain/loss in Table.~\ref{table:baseline_results}). The consistent performance shows that the Cosmic-CoNN trains a more generic and robust CR detector.}}
    \label{fig:gemini_roc}
\end{figure*}

% \tabletypesize{\footnotesize}
\begin{deluxetable*}{ccccc}[t!]
\tablecaption{We evaluated three CR detection methods on LCO imaging data (\S\ref{subsec:ground_results}). The two deep-learning models deepCR and Cosmic-CoNN trained on LCO images are further evaluated on new images from previously unseen Gemini GMOS-N/S telescopes (\S\ref{subsec:unseen_instruments}), with their relative performance loss on the new instruments indicated in \textit{italic parentheses}. Corresponding to the Precision-Recall curves in Fig.~\ref{fig:lco_roc_pr}c and Fig.\ref{fig:gemini_roc}bc, the Cosmic-CoNN model has little or no performance loss, making it a more generic CR detector for new instruments.}
\label{table:baseline_results}
\tablehead{\colhead{Method} & \colhead{Test Data} & \colhead{Precision (\%) at 95\% Recall} & \colhead{TPR (\%) at 0.01\% FPR} & \colhead{TPR (\%) at 0.1\% FPR} \\
 & \mulcccit{(Precision loss/gain on unseen Gemini data)} & }
\startdata
    \hline
    \mulrccc{Astro-SCRAPPY} & LCO Imaging (0m4) & -- & 76.04 & 85.17 \\
    & LCO Imaging (1m0) & -- & 95.03 & 96.41 \\
    & LCO Imaging (2m0) & -- & 99.21 & 99.56 \\
    \hline
    \mulrccc{deepCR\\(LCO-trained)} & LCO Imaging & 89.46 & 99.65 & 99.97 \\
    & GMOS-N/S (1$\times$1 binning) & 76.27 \textit{(-13.19)} & 85.49 & 99.43\\
    & GMOS-S (2$\times$2 binning) & 81.87 \textit{(-7.59)} & 87.58 & 98.97\\
    \hline
    \mulrccc{Cosmic-CoNN\\(LCO-trained)} & LCO Imaging & \textbf{93.70} & \textbf{99.91} & \textbf{99.99} \\
    & GMOS-N/S (1$\times$1 binning) & \textbf{92.03 \textit{(-1.67)}} & 96.40 & 99.84 \\
    & GMOS-S (2$\times$2 binning) & \textbf{96.69 \textit{(+2.99)}} & 97.60 & 99.89 \\
    \enddata
\end{deluxetable*}
% \tablecomments{All values are in (\%).}

% move NRES results out of the table and present in numbers only. 
% \hline
% \hline
% \multirow{1}{*}{\parbox{4cm}{\centering LCO Spectroscopic Data}} & Cosmic-CoNN & 94.4\% & 97.40\% & & 99.86\% \\

% Models used for evaluation table:

% LCO
% 2021_03_14_16_42_LCO_deepCR_continue: 5370 \\
% 2021_03_14_16_36_LCO_Cosmic-Connn_1e3continue: 10040 \\

% HST
% deepCR version 0.1.5
% 2021_03_14_16_42_LCO_deepCR_continue: 5370, 840 complete test set, random rotation \\

% NRS
% 2021_05_05_00_25_NRES_Cosmic-Connn_GN: 7760\\

\section{Results} \label{sec:results}

\edit1{We trained and evaluated the Cosmic-CoNN framework on various types of instruments and data to access its generalization capabilities. Most importantly, we evaluated the \textit{LCO}-trained model on new imaging data from \textit{Gemini Observatory's GMOS-North/South} telescopes \citep{1996RMxAC...4...75G} to understand how well the model generalize to other unseen ground-based instruments. The results are presented in the following structure:}

% \vspace{-4mm}
\begin{itemize}
\setlength\itemsep{-0.5em}
\item Ground-based imaging data
\vspace{-2mm}
\begin{itemize}
    \setlength\itemsep{-0.1em}
    \item Training and evaluation on LCO data (\S4.1)
    \item Evaluating LCO-trained models on Gemini GMOS-North/South data (\S4.2)
\end{itemize}
\item Space-based imaging data (\S4.3)
\item Ground-based spectroscopic data (\S4.4)
\end{itemize}

We \edit1{first} use receiver operating characteristic (ROC) curves as an evaluation metric to compare different detectors' performance at varying thresholds. A ROC curve depicts relative tradeoffs between benefits (true-positive rate, TPR) and costs (false-positive rate, FPR) \citep{DBLP:journals/prl/Fawcett06}. In the context of CR detection:
\begin{align}
    \label{eq:roc_tpr}
    \text{TPR} & = \frac{\text{CR pixels correctly found}}{\text{All CR pixels}} \\
    \label{eq:roc_fpr}
    \text{FPR} & = \frac{\text{Non-CR pixels mistaken as CR}}{\text{All non-CR pixels}}.
\end{align}

Simply put, a higher TPR is desirable at a fixed FPR. While ROC provides a model-wide evaluation at all possible thresholds, standard ROC can be misleading \edit1{for datasets that feature different CR rates (e.g., space- vs. ground-based data). Thus it is not suitable to directly compare a model's TPR given the same FPR between different instruments.}

The Precision-Recall curve, on the other hand, is a more robust metric for imbalanced datasets \citep{2015PLoSO..1018432S}. While recall is equivalent to TPR, in the context of CR detection, precision is defined as:
\begin{align}
    \label{eq:pr_precision}
    \text{Precision} & = \frac{\text{CR pixels correctly found}}{\text{All CR pixels predicted by model}}.
\end{align}

Unlike FPR, precision is determined by the proportion of correct CR predictions given by the model, which is less sensitive to the ratio between CR and non-CR pixels in an image, i.e., it is also less sensitive to the varying CR rates between different datasets. Given a fixed proportion of real CRs correctly discovered (e.g., 95\% recall), the better model should make less mistakes, thus a higher precision. It also helps us to understand how well a model performs on two different datasets given the same recall, or vice versa.

The Precision-Recall curve can also be used as an indicator of prediction confidence. We used this property to provide supplementary evidence that helped \cite{2021NatAs...5..903H} determine a candidate progenitor to be a new type of stellar explosion -- an electron-capture supernova. We rule out the presence of cosmic-ray hits at or around the progenitor site to determine the peak pixel is an actual stellar PSF with $> 3\sigma$ confidence by plotting deepCR's \citep{2020ApJ...889...24Z} predicted score on the corresponding Precision-Recall curve.

\subsection{\edit1{Training and evaluation on LCO data}\label{subsec:ground_results}}

For ground-based imaging data, we randomly sampled and withheld $\sim$$10\%$ of images from the LCO CR dataset as the test dataset. We first analyzed the testset using the filtering-based CR detector Astro-SCRAPPY \citep{curtis_mccully_2018_1482019} for reference. We used \texttt{objlim=2.0} for \textit{LOC} 1.0- and 2.0-meter telescopes' data and \texttt{objlim=0.5} for 0.4-meter for optimal performance in different telescope classes. \texttt{sigfrac=0.1} is held constant for all telescope classes and we produce the ROC curves by varying the \texttt{sigclip} between $\left[1, 20\right]$. Both Cosmic-CoNN and deepCR \citep{2020ApJ...889...24Z} models are trained with identical data and settings. They are evaluated by varying the threshold $t$. Details of the training environment and experiment settings are presented in Appendix~\ref{sec:appendix_training}. 

The Cosmic-CoNN model achieves $99.91\%$~TPR at a fixed FPR of $0.01\%$, outperforming other methods, as illustrated in Fig.~\ref{fig:lco_roc_pr}a,b. \edit1{The Precision-Recall curves in Fig.~\ref{fig:lco_roc_pr}c shows} for both deep-learning models to discover $95\%$ of the real CR pixels ($95\%$ recall), the predictions given by Cosmic-CoNN is over $4\%$ more accurate than deepCR's ($93.70\%$ vs. $89.46\%$ in Precision). If we continue to lower the threshold to allow $99\%$ of the CR pixels being found, Cosmic-CoNN's lead increases to $\sim$$11\%$. Quantitative results are presented in Table~\ref{table:baseline_results}. 

% Cosmic-CoNN's lead seems modest in number but is more obvious when visualized. 
% We analyze the detection discrepancy between different methods and present a few typical examples in Fig.~\ref{fig:LCO_detection}. Binary masks are acquired using the theoretical best threshold of 0.5 for both deep-learning models. We find that Cosmic-CoNN detects CRs more completely, especially the peripheral pixels that other methods tend to miss. The high-fidelity predictions allow Cosmic-CoNN to achieve higher accuracy.

% \begin{figure}[t!]
%     \centering
%     \includegraphics[width=\linewidth]{fig8_lco_detection_compare.png}
%     \caption{Detection discrepancy on \textit{LCO} imaging data. Incorrect or missing CR pixels are marked in red. Cosmic-CoNN detects more peripheral CR pixels than other methods. Examples 1, 4, and 5 in this figure show CRs that are close to real sources, indicating a pattern that deepCR is less robust than other methods in such scenario.}
%     \label{fig:LCO_detection}
% \end{figure}

\subsection{\edit1{Evaluating LCO-trained models on Gemini GMOS-North/South data}} \label{subsec:unseen_instruments}

The goal of this work is to produce a generic ground-based CR detection model. In order to understand how well the models trained on LCO CR dataset perform on unseen instruments, we produced a test dataset consisting of $98$ images from the \textit{Gemini Observatory's GMOS North and South} telescopes \citep{1996RMxAC...4...75G}. The ground-truth CR masks are reduced by the \texttt{DRAGONS} software \citep{2019ASPC..523..321L} with \texttt{hsigma=5.0} to match the setting we used to produce the LCO training data. 

% It is not fair to directly compare the ROC metrics between different datasets, but we could compare two models' relative loss in performance on a new dataset to learn which model is more generic. 

% The two LCO-trained models are tested on new images from \textit{GMOS-N/S} to compare their relative loss in performance, presented . 

\edit1{As shown in Fig.~\ref{fig:gemini_roc} and Table~\ref{table:baseline_results}, at 95\% Recall the deepCR-trained model has $-13.19\%$ and $-7.59\%$ loss in Precision on Gemini's 1$\times$1 and 2$\times$2 binning images, respectively, comparing to its performance on LCO images, while the Cosmic-CoNN model has consistent precisions of $-1.67\%$ and $+2.99\%$. It shows that the Cosmic-CoNN framework is superior in producing more generic models for unseen instruments not included in the training data.}

\begin{figure}[t!]
     \centering
     \includegraphics[trim={0 0 0 20},clip,width=\linewidth]{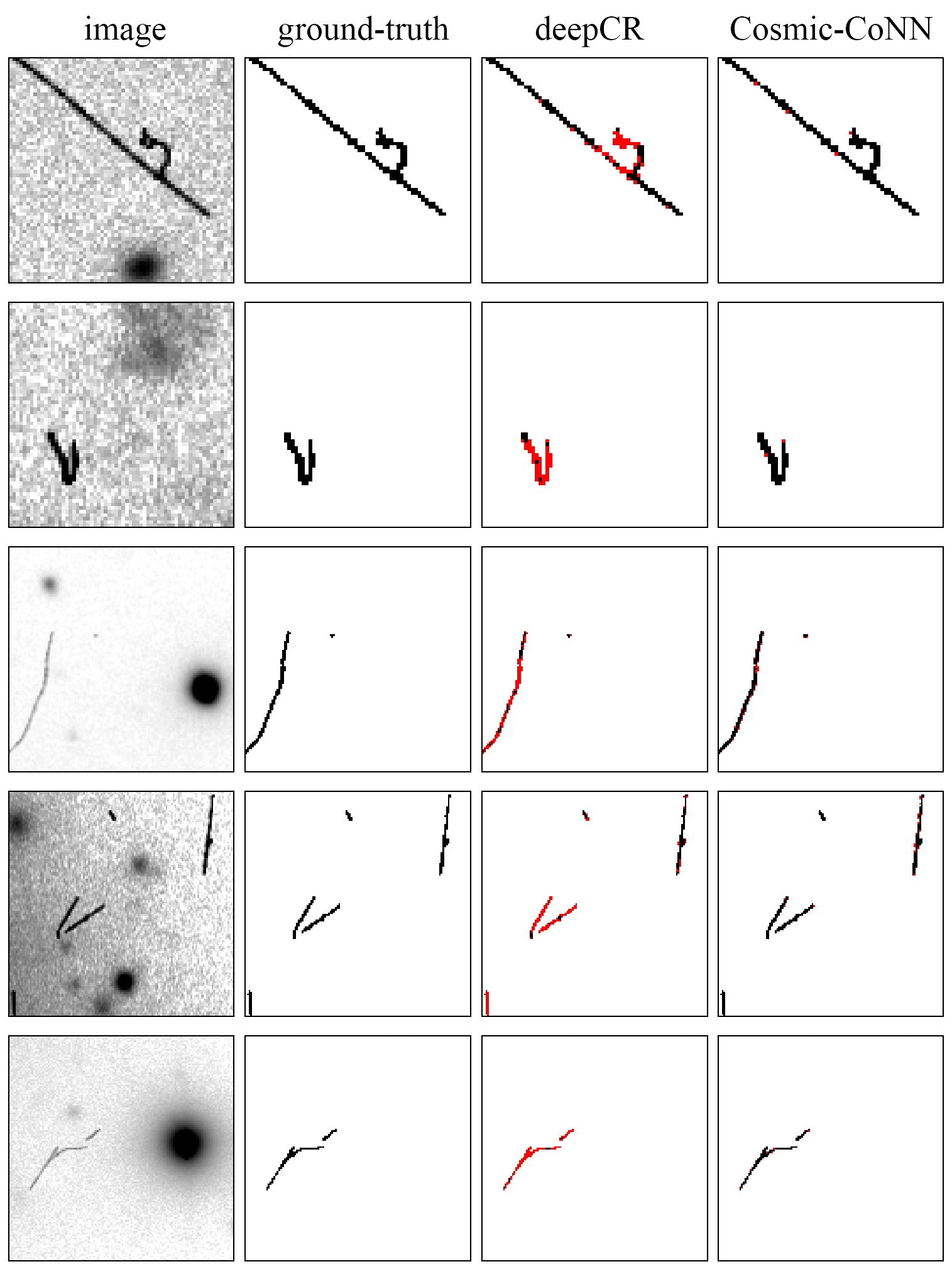}
     \caption{\edit1{deepCR and Cosmic-CoNN were both trained on \textit{LCO} data and tested on \textit{GMOS-N/S} images that were never used for training. While they perform comparably on most CRs, we illustrate some examples that caused deepCR's performance loss on Gemini images (deepCR's 76.27\% \& 81.87\% vs. Cosmic-CoNN's 92.03\% \& 97.69\% in Table~\ref{table:baseline_results}). Both models used the theoretical best threshold of 0.5 for binary masks. Incorrect or missing CR pixels are marked in red.}}
     \label{fig:gemini_detection}
\end{figure}

Examples of detection discrepancy are shown in Fig.~\ref{fig:gemini_detection}. The Cosmic-CoNN model is better at detecting complete CRs of arbitrary shapes, especially the ``worm-shaped'' CRs that frequently appear in the \textit{GMOS-N/S} images.

% The instrument-specific model produced by the deepCR framework performs much better on the test portion of the \textit{LCO} training dataset but lacks the generality for new instruments that are not included in the training data.

\begin{figure*}[t!]
    \centering
    \includegraphics[trim={9 5 5 7},clip,width=\linewidth]{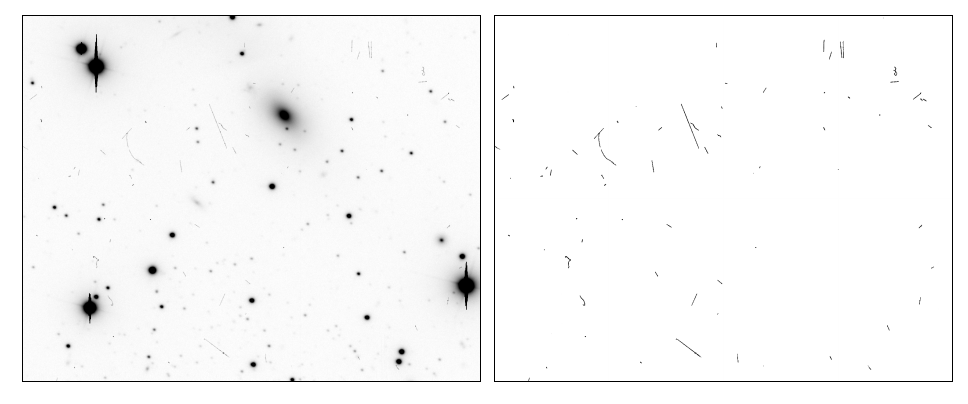}
    \includegraphics[trim={9 9 5 7},clip,width=\linewidth]{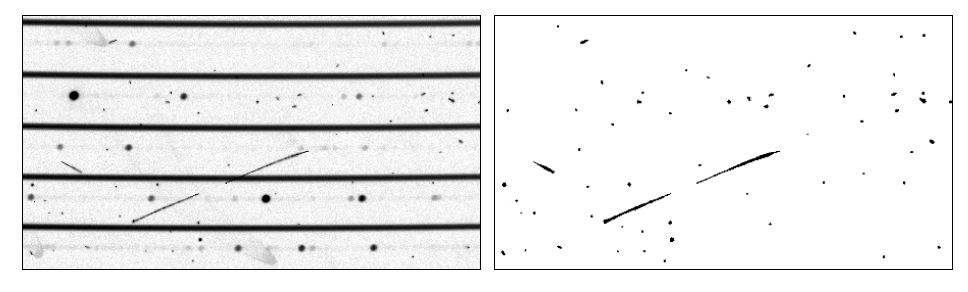}
    \caption{\edit1{A pair of CR detection examples that shows both the Cosmic-CoNN model's generality and the framework's broad applicability. \textbf{(Top)} Cosmic-CoNN's generic ground-imaging model was trained entirely on LCO data yet all visible CRs in a new {Gemini GMOS-N} 1$\times$1 binning image stamp are correctly detected regardless of their shapes or sizes. \textbf{(Bottom)} The Cosmic-CoNN framework also trains well on spectroscopic images and detects CRs over the spectrum robustly on a \textit{LCO NRES} image. The horizontal bands in the left image are the spectroscopic orders, which are left out of the CR mask.}}
    \label{fig:nres_demo}
\end{figure*}

The Cosmic-CoNN model's consistent performance on other CCD imagers also shows the large, diverse LCO CR dataset produces rich cosmic-ray feature coverage that could be effectively generalized to other ground-based instruments. Fig.~\ref{fig:nres_demo} (top row) shows the robust detection result of a heavily CR-contaminated image \edit1{from Gemini GMOS-N}.

% \begin{figure*}[t!]
%     \centering
%     \includegraphics[width=0.95\textwidth]{fig10_HST_ROC_figures.pdf}
%     \caption{\edit1{Despite their similar performance on \textit{HST ACS/WFC} F606W test images (dotted lines), when random mirroring and rotation are applied to the same images to evaluate model robustness (dashdot lines), we notice visible performance loss in the deepCR model, especially in Resolved Galaxy data. Quantatitive results can be found in Table.~\ref{table:hst_results}.}}
%     \label{fig:hst_roc}
% \end{figure*}

%  in extragalactic field and globular cluster observations. But in augmented resolved galaxy images, we see a visible  of $\sim$ $8\%$ TPR at $0.01\%$ FPR in deepCR while Cosmic-CoNN maintains its performance. It shows our improvements designed for ground-based imaging data also applies to training more robust and generic CR detection models for space-based images.

\edit1{\cite{BHAVANAM2022100625} recently tested Cosmic-CoNN on DECam data \citep{Flaugher_2015} and showed it generalizes well to yet another unseen instrument - our Cosmic-CoNN model trained on the LCO CR dataset achieved a precision of 96.60\% at 95.0\% recall, a similar performance as in Gemini's 2$\times$2 binning images (Table \ref{table:baseline_results}). Their improvement of adding attention gate modules \citep{oktay2018attention} only brought marginal performance gain: 0.12\% higher in true positive rate at 0.01\% false positive rate and 0.07\% higher in precision at 95.0\% recall than training with the original Cosmic-CoNN framework. We argue potentially better performance from Cosmic-CoNN as \cite{BHAVANAM2022100625} incorrectly trained on $256^2$ pixel patches, which is against our training strategy discussed in Sec.~\ref{subsec:sampling}.}

\newpage
\subsection{Space-based imaging data}\label{subsec:space_data}

% \begin{deluxetable}{cccc}[t!]
% \tablecaption{123}
% \label{table:hst_results}
% \tablehead{1 & 2 & 3 & 4}
% \startdata
%     \mulrcc{EF} & deepCR & 79.5 \textit{(-0.2)} & 88.6 \textit{(-0.2)} \\
%     & Cosmic-CoNN & 80.2 \textit{(-0.1)} & 89.0 \textit{(-0.1)} \\
%     \hline
%     \mulrcc{GC} & deepCR & 85.4 \textit{(-0.6)} & 93.4 \textit{(-0.3)} \\
%     & Cosmic-CoNN & 86.0 \textit{(0.0)} & 93.8 \textit{(0.0)} \\
%     \hline
%     \mulrcc{RG} & deepCR & 62.1 \textbf{\textit{(-6.8)}} & 75.2 \textbf{\textit{(-6.1)}} \\
%     & Cosmic-CoNN & 63.6 \textbf{\textit{(0.0)}} & 76.3 \textbf{\textit{(0.0)}} \\
% \enddata
% \tablecomments{All values are FPR (\%) in fixed TPR. EF: extragalactic field, GC: globular cluster, RG: resolved galaxy}
% \vspace{-6mm}
% \end{deluxetable}

\begin{deluxetable}{cccc}[t!]
\tablecaption{We reproduced deepCR's results on \textit{HST ACS/WFC} images to compare with Cosmic-CoNN. To test model robustness, we randomly rotated and mirrored the images and indicated each method's performance loss in \textit{italic parentheses}.}
\label{table:hst_results}
\tablehead{\colhead{Data} & \colhead{\hspace{.4cm}Method}\hspace{.4cm} & \colhead{\hspace{.5cm}0.01\% FPR}\hspace{.5cm} & \colhead{0.05\% FPR} \\ & \mulcccit{(TPR loss w/ mirror+rotation)}}
\startdata
    \mulrcc{EF} & deepCR & 79.5 \textit{(-0.2)} & 88.6 \textit{(-0.2)} \\
    & Cosmic-CoNN & 80.2 \textit{(-0.1)} & 89.0 \textit{(-0.1)} \\
    \hline
    \mulrcc{GC} & deepCR & 85.4 \textit{(-0.6)} & 93.4 \textit{(-0.3)} \\
    & Cosmic-CoNN & 86.0 \textit{(0.0)} & 93.8 \textit{(0.0)} \\
    \hline
    \mulrcc{RG} & deepCR & 62.1 \textbf{\textit{(-6.8)}} & 75.2 \textbf{\textit{(-6.1)}} \\
    & Cosmic-CoNN & 63.6 \textbf{\textit{(0.0)}} & 76.3 \textbf{\textit{(0.0)}} \\
\enddata
\tablecomments{All values are FPR (\%) in fixed TPR. EF: extragalactic field, GC: globular cluster, RG: resolved galaxy}
\vspace{-6mm}
\end{deluxetable}

% For space-based imaging data, we trained the Cosmic-CoNN model on the \textit{HST ACS/WFC} dataset that \cite{2020ApJ...889...24Z} released and compared with the official released deepCR CR detection model.

We also trained Cosmic-CoNN on \cite{2020ApJ...889...24Z}'s \textit{HST ACS/WFC} F606W dataset consisting of extragalactic field, globular cluster, and resolved galaxy observations to demonstrate the framework's broad applicability. The Cosmic-CoNN-trained model has better performance in all three types of observations comparing to the deepCR model (version 0.1.5), \edit1{as shown in Table.~\ref{table:hst_results}.} When testing model robustness on augmented images with random mirroring and rotation \citep{2018A&C....25..103G}, we found more robust performance from Cosmic-CoNN with little or no performance loss, especially in resolved galaxy data.

Unlike the LCO CR dataset which releases full-size images in \texttt{FITS} format, the F606W dataset sliced and stored images as $256^2$ pixel stamps in \texttt{Numpy} arrays, so we were not able to test the effect of increased sampling stamp size on these data. \cite{Kwon_2021} recently trained an all-filter \textit{HST ACS/WFC} deepCR model on an extended dataset covering the entire spectral range of the ACS optical channel. Cosmic-CoNN supports loading deepCR models to use with our toolkit, instructions are available at \href{https://github.com/cy-xu/cosmic-conn}{\tt https://github.com/cy-xu/cosmic-conn}.

\subsection{Ground-based spectroscopic data}\label{subsec:spectroscopic_data}

% Finally, we trained a Cosmic-CoNN spectroscopic model using \textit{LCO} data and saw exceptional performance (\S\ref{subsec:spectroscopic_data}). 

% considering the high CR rates in images of 15 minutes or longer exposure time.

% colhead{Method} & \colhead{Data} & \colhead{Precision at 95\% Recall} & \colhead{TPR at 0.01\% FPR} & & \colhead{TPR at 0.1\% FPR} \\
% LCO Spectroscopic Data & Cosmic-CoNN & 94.4\% & 97.40\% & & 99.86\% \\

Finally, we expand the Cosmic-CoNN framework to detecting CRs in single-exposure spectroscopic images, a task that has remained challenging for conventional methods. \cite{2017PASP..129b4004B} was able to detect as many as $80\%$ of the CRs in single-exposure, multi-fiber spectral images. Based on two-dimensional profile fitting of the spectral aperture, their method takes about 20 minutes to process a $4K\times4K$ pixel image. Cosmic-CoNN detects nearly all CRs in about 25 seconds on CPU and less than 5 seconds with GPU acceleration.

To prepare the data for deep-learning training, we modified our custom CR-labeling pipeline (Appendix \ref{appendix:cr_pipeline}) and produced a dataset of over $1,500$ images using repeated observations from the four instruments of \textit{LCO}'s \textit{Network of Robotic Echelle Spectrographs} (\textit{NRES}) located around the world. We randomly sampled and reserved $20\%$ of the data as the test set and used the rest for training and validation.

Cosmic-CoNN reaches $97.40\%$ TPR at $0.01\%$ FPR with a precision of $94.4\%$ at $95\%$ recall. Considering the high CR rates in spectroscopic images because of the 15 minutes or longer exposure time, the NRES model in fact demonstrates exceptional performance. A detection result example is shown in Fig.~\ref{fig:nres_demo} (bottom row). We consider these results preliminary because the focus of this paper is on a generic ground-based imaging model and we will conduct thorough comparison with other methods in a future work. Nevertheless, the versatility of Cosmic-CoNN framework potentially paves a way for solving the CR detection problem in the accuracy-demanding spectroscopic data.

% \edit1{The Cosmic-CoNN NRES model demonstrates the exceptional performance of} $97.40\%$ TPR at $0.01\%$ FPR with a precision of $94.4\%$ at $95\%$ recall, considering the high CR rates of LCO spectroscopic images for their 15 minutes or longer exposure time.

% \cite{2017PASP..129b4004B} Multi-fiber images do not have clear isolated point or extended sources as in the photometric data, and the long stripe-like multi-fiber spectra occupy large contiguous regions so that the available area for the local "background" is much smaller than in the photometric data. Methods with median filtering or interpolation of neighboring pixels are less effective in this case.

\section{Toolkit} \label{sec:toolkit}

We have built a suite of tools to democratize deep-learning models in order to make automatic, robust, and rapid CR detection widely accessible to astronomers. The toolkit includes console commands for batch processing \texttt{FITS} files, a web-based app providing CR mask preview and editing capabilities, and Python APIs to integrate Cosmic-CoNN models into other data workflows. 

The Python toolkit package is released on \texttt{PyPI}. We host the open-source Cosmic-CoNN framework on GitHub \href{https://github.com/cy-xu/cosmic-conn}{\tt https://github.com/cy-xu/cosmic-conn} with complete documentation including toolkit manual, developer instructions on using the LCO CR dataset and training new models. We also released the LCO CR dataset and the code used to generate the results to facilitate reproducibility.

\begin{figure}[t!]
    \centering
    \frame{\includegraphics[clip,width=\linewidth]{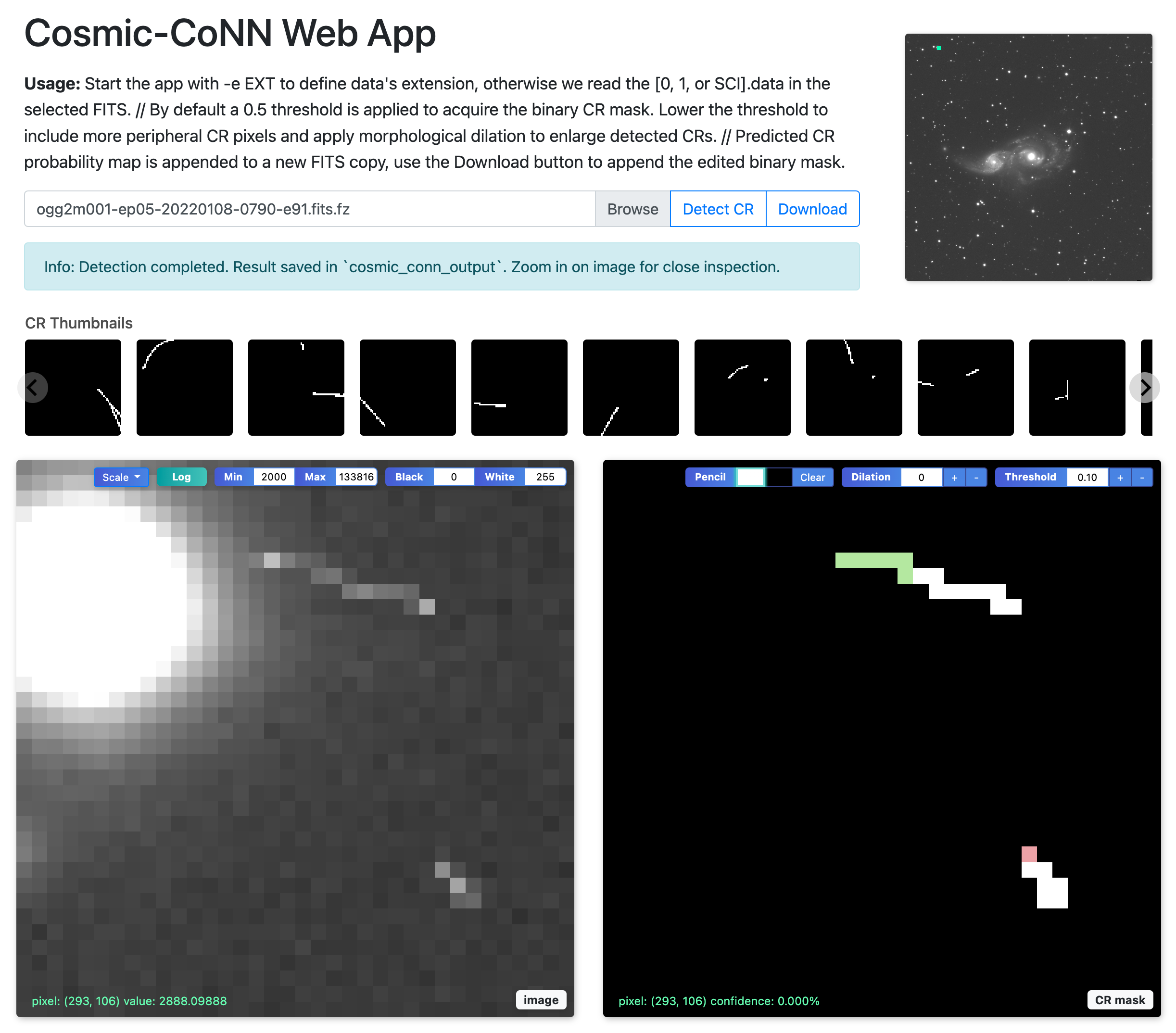}}
    \caption{The \edit1{interactive web app streamlines the workflow of CR detection, visualization, and mask editing into a single interface \citep{Xu_2022_CVPR}.} This tool is also useful to help users find the suitable threshold for new data. Users could adjust the threshold, apply morphological dilation, or perform pixel-level manual editing on the CR mask to acquire the desirable results for downstream analysis.}
    \label{fig:web_app}
\end{figure}

Console commands are the most convenient way to perform batch CR detection on \texttt{FITS} files directly, e.g.,\\ \texttt{\$ {cosmic-conn -i input -m ground\_imaging}} \\
utilizes the generic \verb|ground_imaging| model and the user can replace the argument with \verb|NRES| or the path to a new model trained with Cosmic-CoNN for other types of data. The result is attached as a \texttt{FITS} extension. In terms of speed, Cosmic-CoNN provides more accurate prediction than conventional methods in comparable time on the CPU. Processing a $2K\times 2K$ pixels image takes $\sim$7.5s on a \texttt{AMD Ryzen 9 5900HS} laptop processor. With GPU-acceleration, it takes only $\sim$0.8s on a high-end \texttt{Nvidia Tesla V100} GPU, and $\sim$1.2s on an entry-level \texttt{Nvidia GTX 1650} laptop GPU.

The \texttt{\$ cosmic-conn -a} command starts an interactive CR detector in the browser, as shown in Fig.~\ref{fig:web_app}. We adopt the interface layout and controls from the \texttt{SAOImageDS9}  \citep{2003ASPC..295..489J}. In addition, we provide an array of CR thumbnails for quick navigation and the ability to edit CR masks in real time. The \texttt{JavaScript}-backed web app provides necessary tools for users to fine-tune the appropriate post-processing parameters for different instruments. The preview window supports various scaling methods like the \texttt{zscale} for better visualization.

% \vspace{-0.2cm}
\begin{figure}[ht]
    Cosmic-CoNN is designed to be integrated in custom data pipelines. Let \texttt{image} be a two-dimensional \texttt{float32} array: \newline
    {\includegraphics[trim={0 0 0 0},clip,width=\linewidth]{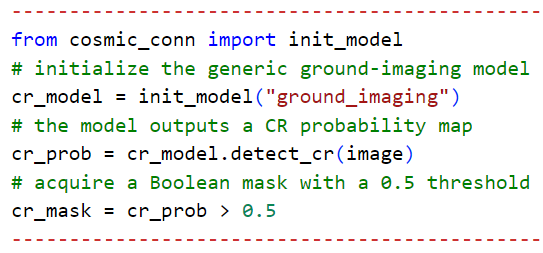}}
\end{figure}
% \vspace{-0.2cm}

Our \texttt{Python} APIs allows other facilities to integrate rapid CR detection into their data reduction pipeline. The framework checks if the host machine supports GPU-acceleration and prioritizes computation on GPU. Then it optimizes the detection strategy (full image or slice-and-stitch using smaller stamps) based on available memory without human intervention. 

We are planning to deploy the web app on the cloud to provide GPU-accelerated CR detection as a free service. This will allow users to upload their failure cases to us to expand the training set and improve the model. In the current release, the web app is a local instance which does not collect or upload any user information.

% \newpage
\section{Conclusion} \label{sec:conclusion}

In this work, we presented an end-to-end solution to help tackle the CR detection problem in astronomical images. The large, diverse LCO CR dataset produces rich feature coverage, allowing deep-learning models to achieve state-of-the-art CR detection on single-exposure images from Las Cumbres Observatory. The Cosmic-CoNN deep-learning framework trained generic CR detection models that maintain \edit1{consistent} performance on unseen instruments. Extensive evaluation showed the framework's broad applicability in ground- and space-based imaging data, as well as spectroscopic data. Finally, we released a toolkit to make the deep-learning CR detection easily accessible to astronomers.

Using the generic Cosmic-CoNN model as a pre-trained initialization, other facilities could fine-tune a model optimized for their own CCD imager with a lot less data. The LCO CR dataset also lays the foundation for a potential universal solution. By expanding our dataset with more instruments from other facilities, we are confident to see an universal CR detection model that achieves better performance on unseen ground-based instruments without further training.

The Cosmic-CoNN framework and the toolkit will be a valuable resource for the community to develop future deep-learning methods for source extraction, satellite detection, near-Earth objects detection, and more. These topics are not the focus of this paper but our improvements to the neural network made Cosmic-CoNN a suitable deep-learning architecture for these tasks, as we have seen in some preliminary experiments.

With the current Cosmic-CoNN model rejecting CRs that could be falsely recognized as astronomical sources, we could better profile the point spread functions in order to address the $\sim$1.21\% excluded CR pixels in the next release of our dataset. We expect to see further improvement in the Cosmic-CoNN model.

As large surveys like the \textit{Vera Rubin Observatory}'s \textit{Legacy Survey of Space and Time} (\textit{LSST}) \citep{2019ApJ...873..111I} go online, we will see an explosion of new data that requires automatic, robust, and rapid CR detection. With GPU-acceleration, deep-learning methods like Cosmic-CoNN will likely be the solution for future data reduction pipelines that is needed to process the over 100 terabytes of data produced each night from \textit{LSST} and many follow-up facilities.

\acknowledgments

We thank Yuxiang Wang, Jiaxiang Jiang, Chris Hellmuth, Keming Zhang, Jennifer Jacobs, and Tobias H\"{o}llerer for their discussion and feedback on this work. We thank Simon Conseil and the \texttt{DRAGONS} software \citep{2019ASPC..523..321L} support team for their help in producing the \textit{Gemini GMOS} evaluation dataset.

This work makes use of observations from the \textit{Las Cumbres Observatory} global telescope network \citep{Brown_2013}. This works is also based on observations obtained at the international Gemini Observatory, a program of NSF's NOIRLab, which is managed by the Association of Universities for Research in Astronomy (AURA) under a cooperative agreement with the National Science Foundation on behalf of the Gemini Observatory partnership: the National Science Foundation (United States), National Research Council (Canada), Agencia Nacional de Investigaci\'{o}n y Desarrollo (Chile), Ministerio de Ciencia, Tecnolog\'{i}a e Innovaci\'{o}n (Argentina), Minist\'{e}rio da Ci\^{e}ncia, Tecnologia, Inova\c{c}\~{o}es e Comunica\c{c}\~{o}es (Brazil), and Korea Astronomy and Space Science Institute (Republic of Korea).

Use was made of computational facilities purchased with funds from the National Science Foundation (OAC-1925717) and administered by the Center for Scientific Computing (CSC). The CSC is supported by the California NanoSystems Institute and the Materials Research Science and Engineering Center (MRSEC; NSF DMR 1720256) at UC Santa Barbara. This work was also
partially funded by National Science Foundation grants IIS-1619376 and IIS-1911230.


\facilities{\textit{LCOGT}, \textit{HST(ACS/WFC)}, \textit{Gemini:Gillett}, \textit{Gemini:South}}


\software{\texttt{Astropy} \citep{astropy:2013, astropy:2018}, \texttt{Astro-SCRAPPY} \citep{curtis_mccully_2018_1482019}, \texttt{Cosmic-CoNN} \citep{cy_xu_2022_6630624, 2021ascl.soft08018X}, \texttt{DRAGONS} \citep{2019ASPC..523..321L}, \texttt{reproject} \citep{2020ascl.soft11023R}, \texttt{Matplotlib} \citep{Hunter:2007}, \texttt{NumPy} \citep{harris2020array}, \texttt{scikit-image} \citep{scikit-image},  \texttt{SExtractor} \citep{1996A&AS..117..393B}, \texttt{PyTorch} \citep{2019arXiv191201703P}}

% this is a temporary layout adjustment
\clearpage



\appendix

\section{CR Labeling Pipeline\label{appendix:cr_pipeline}}

The ground-truth CR-labeling pipeline starts with searching for successive exposures of the same field. We acquire the publicly available scientific observations from \textit{LCO}'s Science Archive\footnote{https://archive.lco.global/} and filter the number of visits users requested (more than three but no more than twelve). It is unlikely a cosmic ray will hit the same pixel location twice, so every three consecutive exposures are saved as a sequence into a multi-extension \texttt{FITS} file for alignment and CR labeling, while maintaining all the header information for future community research. For higher signal-to-noise ratio and higher CR rates, we only used images with an exposure time of 100 seconds or longer. We further constrained the consecutive images to be taken within the same schedule molecule, the minimal \textit{LCO} scheduler unit. Images from the same molecule ensure intervals between exposures are minutes or less, which minimize the variations in seeing conditions and point spread function (PSF). We reject a sequence whose background varies over $\sigma > 5$ between frames, as they are not stable enough to robustly identify cosmic rays.

We then reproject to align each frame in the sequence with \texttt{astropy/reproject} \citep{robitaille_ginsburg_deil_2019} using nearest-neighbor interpolation to ensure CRs are not distorted during re-sampling. Fig.~\ref{fig:cr_triplets} shows an image stamp from an aligned sequence. \textit{LCO}'s \texttt{BANZAI} \citep{2018SPIE10707E..0KM} data reduction pipeline have bias and dark frame subtracted to remove instrument signature, allowing us to use one CR-labeling pipeline across all \textit{LCO} instruments. Let $I$ be an image in the sequence then $I$'s noise uncertainty $\sigma_I$ is simplified to:
\begin{equation}
\sigma_I = \sqrt{\left| I \right| + N_{R}^2 + N_{S}}
\end{equation}
where $N_R$ is the CCD read noise, $N_{S}$ is the sky background noise, which corrects for the background variation between exposures. We then approximate the median frame uncertainty $\Sigma$ by performing median filtering at each pixel location across the uncertainties from the three frames $I_1$, $I_2$, and $I_3$ in order to reject the variance from the CR pixels:

\begin{equation}
\Sigma = \frac{Median(\sigma_{I_1}, \sigma_{I_2} \sigma_{I_3})}{\sqrt{3}}.
\end{equation}

We update each frame $I$ with sky subtraction $I := I - Median(I)$ before calculating the median frame $M_{I}$. We then define a deviation score that calculates how much each frame deviates from the median frame represented in Gaussian distribution:

\begin{equation}
\text{Deviation score} = \frac{I - M_I}{\sqrt{(\sigma_I)^2 + \Sigma^2}}.
\end{equation}

Pixel locations with a deviation score $>5.0$ are identified as bright CR pixels and labeled in a preliminary outlier mask. A morphological dilation of five pixels is applied to the outlier mask, and we use a lower threshold of $>2.5$ to include the dimmer peripheral pixels around the CRs.

A key step to acquire the final CR mask is to remove false-positive outliers caused by PSF wings and isolated hot pixels. We perform source extraction with \texttt{SEP} \citep{barbary_2016} on the CR-free median frame to acquire a robust source catalog. We then perform windowed background estimation to include the astrophysical source pixels in an ignore mask to reject false-positive outlier from PSF wings \citep{howell_2006_photometry}.

\texttt{BANZAI} provided a mask for permanent dead CCD pixels but we also noticed a very small fraction of remaining standalone hot pixels that are more likely to be Poisson noise or persistent pixels due to over saturation in previous exposures. Thus our last step is to reject isolated (single) hot pixel events to acquire the final CR mask. Different types of artifacts and rejected pixels, including 100 pixels ignored around CCD boundaries are coded and included in the ignore mask. Instruction on using the data pipeline, the LCO CR dataset, and the ignore mask coding rules can be found in the documentation \href{https://github.com/cy-xu/cosmic-conn}{\tt https://github.com/cy-xu/cosmic-conn}.

\section{Ablation Study\label{sec:ablation_study}}

An ablation study helps us understand how a building block or a design choice affects a machine learning system's overall performance. It applies or removes a single component in a controlled experiment while holding other parameters constant. We evaluate the proposed improvements discussed in Sec.~\ref{sec:dl_framework} through variant models corresponding to Fig.~\ref{fig:training_evaluation} and present the quantitative results in Table.~\ref{table:ablation}.

% Less epochs means a less convoluted convergence path, as shown in Fig.~\ref{fig:training_evaluation}. 

\begin{deluxetable}{lccccccc}[h!]
    % \tablenum{1}
    % \tabletypesize{\footnotesize}
    \tablecaption{Cosmic-CoNN ablation study on \textit{LCO} and \textit{Gemini} imaging data. All variant models are evaluated with identical validation images and the same input stamp size. We gauge training efficiency by the number of \texttt{epochs} a model takes to reach a Dice score $>0.85$ \citep{sorensen1948method} during training, corresponding to convergence curves in Fig.~\ref{fig:training_evaluation}. We discussed in Sec.~\ref{sec:results} that Precision is less sensitive to the varying CR rates between different datasets than TPR at fixed FPR, thus we measure a model's Precision at 95\% Recall on \textit{LCO} and \textit{Gemini} data to evaluate how well it generalizes to unseen data, corresponding to a model's performance at epoch 4000 shown in Fig.~\ref{fig:training_evaluation}, higher is better.}  \label{table:ablation}
    \tablehead{
    \colhead{Method} & \colhead{Dice score $>0.85$} & \colhead{LCO Precision} & \colhead{Gemini 1$\times$1 Precision} & \colhead{Gemini 2$\times$2 Precision}
    }
    \startdata
    deepCR (baseline) & 2980 & 89.19\% & 79.59\% & 84.88\% \\
    deepCR + Median-Weighted loss & 2080 & 92.98\% & 78.76\% & 83.08\% \\
    deepCR + $1024^2$px & n/a & 89.35\% & 82.57\% & 86.55\% \\
    deepCR + GN & 1420 & 90.82\% & 77.07\% & 89.30\% \\
    deepCR + $1024^2$px + GN & 1040 & 93.17\% & 84.54\% & 92.09\% \\
    Cosmic-CoNN (MW loss + $1024^2$px + GN) & \textbf{380} & \textbf{93.40\%} & \textbf{86.80\%} & \textbf{94.37\%} \\
    \enddata
\end{deluxetable}
\vspace{-1cm}
The complete ablation study (combining quantitative results from Table.~\ref{table:ablation} with training visualizations in Fig.~\ref{fig:training_evaluation}) shows applying the proposed Median-Weighted loss function to the baseline method improves model performance on LCO data from $89.19\%$ to $92.98\%$, at the same time improves training efficiency from 2980 to 2080 epochs, which validates that the new loss function does indeed provide a better model convergence path discussed in \S\ref{subsec:loss}.

While the Median-Weighted loss alone does not produce a more generic model, all variant models trained with the larger $1024^2$ pixel sampling stamps demonstrated better model generality on the unseen Gemini data, especially the $1024^2$px + group normalization (GN) combination that we discussed in \S\ref{subsec:sampling}. GN alone does not improve performance but mainly contributes to training efficiency, which is better visualized in Fig.~\ref{fig:training_evaluation} when compared with models that adopt the two-phase training.

The proposed Median-Weighted loss further provided the ($1024^2$px + GN) variant model a better convergence path to produce the Cosmic-CoNN model that excels in both training efficiency (from 2980 to 380 epochs) and performance on not only LCO instruments which were used for training (from $89.19\%$ to $93.40\%$) but also Gemini instruments that were not included in training data (from $79.59\%$ to $86.80\%$ on $1\times1$ binning \& from $84.88\%$ $94.37\%$ on $2\times2$ binning) among all variant models.

The ablation study shows each of our proposed improvements affects certain aspects of the machine learning system and their joint effect contributes to the generic and best-performing Cosmic-CoNN model suitable for the CR-detection task in ground-based astronomical data with variable conditions from multiple instruments.

% a single component improves or hurts model performance. 

% while the Cosmic-CoNN model, which benefits from the joint effects of the novel median-weighted loss, the larger $1024^2$ pixel sampling size, and group normalization (GN) performs the best in both training efficiency and generality.

\section{Training Details}\label{sec:appendix_training}

We implement the Cosmic-CoNN framework in \texttt{PyTorch} 1.6.0 \citep{2019arXiv191201703P} with Adam optimizer \citep{2014arXiv1412.6980K}. Models for the same type of observation are trained with identical data, random seed, and hardware. We use the \texttt{Nvidia Tesla v100 32GB} GPU for training. The large GPU memory allows us to maximize the batch size $n$ in each iteration. All training settings are identical unless it is clearly specified for a variant model. Scripts to reproduce our experiments are included in the source code.

For LCO imaging data, we randomly sampled and withheld $20\%$ of the training set for validation. An initial learning rate of $0.001$ was used for all models. During training, we monitor the validation loss for each model and manually decay the learning rate by $0.1$ when the loss plateaus. In the ablation study, we reduce the learning rate to 0.0001 at epoch 3,000 for all models. Models using group normalization adopt a fixed \texttt{group=8} for all feature layers. For the median-weighted loss we linearly scale the lower bound $\alpha$ from 0 to 1 over 100 epochs. We re-implemented deepCR with identical network and adopted the two-phase training that \cite{2020ApJ...889...24Z} used to train deepCR models. The Cosmic-CoNN batch normalization (BN) variant model also adopted the two-phase training. In order to make fair comparisons, all Cosmic-CoNN and deepCR models were carefully tuned, the best models were used for evaluation.

The Cosmic-CoNN model and variant models with $1024^2$ pixels sampling stamp size used a batch size of $n=10$ in the ablation study. deepCR and its variant models adopt $256^2$ pixels stamp size with $n=160$ to ensure the model sees the same amount of pixels in a mini-batch. For a dataset of $N$ samples, models trained with batch size $n=10$ updates $\lfloor\frac{N}{10}\rfloor$ times in an epoch but models trained with $n=160$ only update $\lfloor\frac{N}{160}\rfloor$ times, which leads to unfair comparisons on training efficiency. We addressed this issue by sampling a subset of $\lfloor\frac{N}{16}\rfloor$ samples as an epoch for models with batch size $n=10$.

For \textit{HST ACS/WFC} imaging data, the Cosmic-CoNN model is trained on identical data as deepCR \citep{2020ApJ...889...24Z} but with a new \texttt{PyTorch} data loader that added random rotation and mirroring while sampling images. The larger GPU memory allowed us to use $256^2$ pixels sampling stamp size with $n=160$.

For \textit{LCO NRES} spectroscopic data, the neural network is identical to the Cosmic-CoNN ground-imaging model. We used a stamp size of $1024^2$ pixels with $n=8$, an initial learning rate $0.0001$, and manually monitor and decay the learning rate.


\bibliography{ms}{}
\bibliographystyle{aasjournal}



\end{CJK*}
\end{document}